# High-frequency dynamics of type-B glass formers investigated by broadband dielectric spectroscopy

S. Kastner[1], M. Köhler[1], Y. Goncharov[2], P. Lunkenheimer[1,*], and A. Loidl[1]

[1]*Experimental Physics V, Center for Electronic Correlations and Magnetism, University of Augsburg, 86135 Augsburg, Germany*
[2]*Institute of General Physics, Russian Academy of Sciences, 119991 Moscow, Russia*

**Abstract**

We present the results of broadband dielectric spectroscopy on two glass formers with strong Johari-Goldstein β-relaxations. In addition to the α- and β-relaxation dynamics, the extension of the spectra up to 1 THz also allows revealing information on the fast β-process in this class of materials. There is clear evidence for a fast process contributing in the region of the high-frequency loss minimum, which is analyzed in terms of the idealized mode-coupling theory.

## 1. Introduction

In recent years the dynamics of glass formers at frequencies beyond that of the structural α-relaxation has attracted considerable interest [1,2,3]. The most prominent of these fast processes is the Johari-Goldstein (JG) β-relaxation [4], also termed slow β-process to distinguish it from the fast β-process predicted by mode-coupling theory (MCT) [5]. While various explanations of this phenomenon have been proposed (see, e.g., [4,6,7,8]), no consensus on its microscopic origin has been reached until now. Glass formers whose spectra of the dielectric loss $\varepsilon''(\nu)$ (or other susceptibilities) exhibit a well-pronounced slow β-relaxation peak are often termed "type B" [9,10]. In contrast, in type A glass formers only a second power law at the high-frequency wing of the α-peak is found, a spectral feature that has been termed "excess wing" [11,12]. In contrast to the slow β-relaxation, the fast β-relaxation (often simply termed "fast process") usually is assumed to show up at much higher frequencies, typically at 10 GHz - 1 THz [5,13]. It does not lead to a separate loss peak in the spectra but generates excess intensity in the region of a shallow minimum found at these frequencies. Dielectric investigations, nearly continuously covering the whole dynamic region from the α- up to the fast β-process are scarce. Previously our group has reported such results for several type A [2,3,14,15] but only for two type B glass formers, namely di- and tripropylene glycol [15] whose secondary relaxations were classified as not being of JG type [16,17]. In the present work, we provide such spectra on the two type B glass formers xylitol and sorbitol, including the frequency regimes of α-, slow β- *and* the fast β-relaxation. In addition to results on the α- and JG β-relaxation, information on the fast β-process and an analysis using idealized MCT is provided.

## 2. Experimental procedures

Xylitol ($T_g$ = 248 K) was purchased from Aldrich with a purity ≥ 98% and sorbitol ($T_g$ = 274 K) from AppliChem with a purity of 99.7%. All samples were used without further purification. A combination of

---

different experimental techniques is necessary to record the real and imaginary part of the dielectric permittivity in the broad frequency range covered by the present work. For the aging and low-frequency measurements at $10^{-4} \leq \nu \leq 3 \times 10^{6}$ Hz, parallel plane capacitors filled with the sample material were measured using a frequency response analyzer (Novocontrol α-analyzer). At higher frequencies, $10^{6} \leq \nu \leq 3 \times 10^{10}$ Hz, a coaxial reflectometric technique was used employing several impedance and network analyzers (Agilent E 4991A, HP 4291B, and HP 8510C). Beyond this range, measurements at frequencies from 60 GHz up to 1.4 THz were carried out using a Mach-Zehnder interferometer [18]. For further details on these techniques the reader is referred to [2,19].

## 3. Results and discussion

### 3.1. α- and slow β-process

Figures 1 and 2 show the broadband dielectric loss spectra of xylitol and sorbitol, respectively. Part of the results on xylitol, but with only sparse data in the minimum region, have been previously published in [20]. Aside of the α-peak, both glass formers show a strong slow β-relaxation, i.e., they are typical type B glass formers. The β-relaxations in both xylitol and sorbitol were classified as "genuine" JG relaxations [21,22]. At high frequencies, similar to the findings in type A glass formers [2,3,14,15], a shallow minimum shows up.

The α- and slow β-relaxation dynamics of both materials have been extensively treated in various previous publications (e.g., [21,22,23,24,25,26,27,28,29,30,31,32]) and thus will be only briefly discussed in the following. For both materials, the typical asymmetric α-peaks are observed whose strong shifting with temperature mirrors the glassy freezing of the structural dynamics. The slow β-peak is well separated from the α-peak at low temperatures and develops into a shoulder with increasing temperature before showing the usual merging with the α-peak [4,30,33]. When approaching the merging region at high temperatures (around 300 K for both materials), the amplitude of the slow β-relaxation reaches similar magnitudes as that of the α-peak. The lines in Figs. 1 and 2 are fits of the frequency region below the minimum with the sum of a Havriliak-Negami function [34] for the α-peak and a Cole-Cole function [35] for the β-peak. For sorbitol at low temperatures an additional term $\varepsilon'' = \sigma_{dc}/(2\pi\nu\varepsilon_0)$ (with $\varepsilon_0$ the permittivity of vacuum) had to be used to take into account a non-negligible conductivity contribution $\sigma_{dc}$ from impurity-generated ionic charge transport. Following common practice, for the sake of clarity in Figs. 1 and 2 this contribution, usually leading to a divergence of $\varepsilon''(\nu)$ at frequencies below the left wing of the α-peaks, is not shown. However, in sorbitol it becomes increasingly important at low temperatures and no longer can be neglected. This obviously arises from a different temperature dependence of α-relaxation time and dc conductivity, implying a strong violation of the Debye-Stokes-Einstein relation [36] in sorbitol.

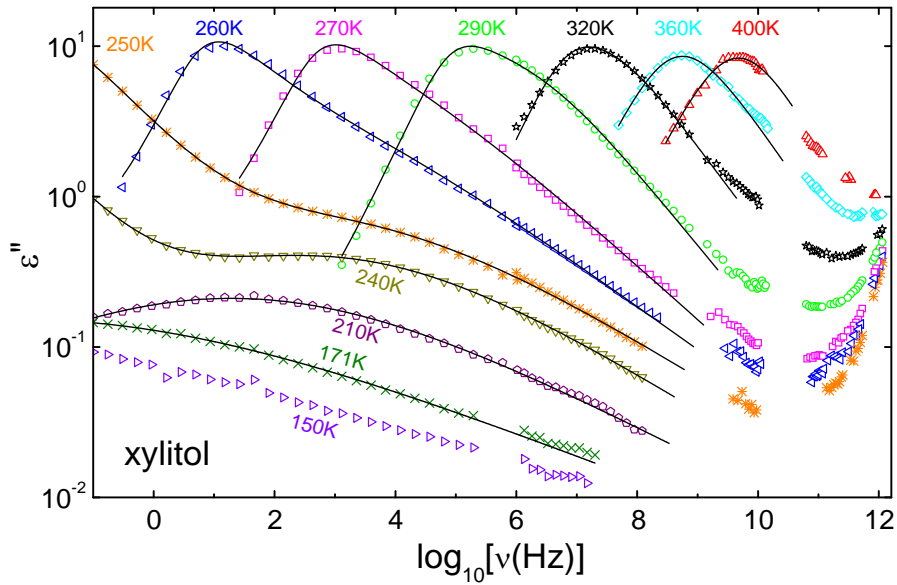

Fig. 1. Broadband dielectric loss spectra of glass forming xylitol at selected temperatures. The lines are fits with the sum of a Havriliak-Negami and a Cole-Cole function. The data below 120 GHz have been previously published in [20].

It should be noted that most of the loss spectra with well developed β-peaks in Figs. 1 and 2 have been measured below the glass temperature and thus were not taken in thermodynamic equilibrium. However, in sorbitol ($T_g$ = 274 K, Fig. 2) the curves at 272, 267, and 261 K were taken after sufficiently long aging times (16 h, 51 h, and 12 days, respectively) to ensure that equilibrium was reached. Interestingly, at around 1 Hz the spectrum for sorbitol at 261 K reveals excess intensity in the region between α-peak (of which only the high-frequency flank is seen) and the β-peak. This spectrum, as well as that at 267 K could only be fitted when assuming an additional third relaxation process. Qualitatively the situation seems to be similar as in 3-flouroaniline [9] and some propylene glycol oligomers [16,37] where a further secondary relaxation or an excess wing was found in the spectra at low temperatures.

Figure 3 shows the temperature dependences of the relaxation times $\tau$ of both materials in Arrhenius representation. For xylitol (Fig. 3(a)) [20], the closed symbol indicates a data point for $\tau_\alpha$ obtained from an evaluation of the time dependence of the loss below $T_g$ [38] providing a reasonable extension of the $\tau_\alpha(T)$ curves from equilibrium measurements. The α-relaxation times show pronounced non-Arrhenius behavior and have been parameterized by the modified Vogel-Fulcher-Tammann relation, $\tau_\alpha \propto \exp[DT_{VF}/(T-T_{VF})]$ with $T_{VF}$ the Vogel-Fulcher temperature and $D$ the strength parameter (lines in Fig. 3) [39,40]. In addition, the β-relaxation times $\tau_\beta$ are included in Fig. 3. It should be noted that only part of these data were taken in equilibrium as indicated by different symbols in the figure. The β-relaxation times $\tau_\beta(T)$ of both materials show Arrhenius characteristics at low temperatures, $T < T_g$, which is typical for secondary relaxations. Above the glass temperature, $\tau_\beta(T)$ seems to cross over into a stronger temperature dependence than Arrhenius as was also found in other materials (e.g., [3,15,27,30,41,42]). At $T > T_g$, the β-relaxation times of both materials closely approach the $\tau_\alpha(T)$ curves. However, one should be aware that in this region there is a high uncertainty in the determination of $\tau_\beta$ due to the strong overlap with the α-relaxation. For xylitol, a minimum in $\tau_\beta(T)$ occurs close to $T_g$ (Fig. 3(a)) and some indication for such a minimum is also found for sorbitol (Fig. 3(b)). The occurrence of minima in the relaxation times of secondary relaxations was also found in other glass formers [15,37,43,44]. It can be understood within the so-called minimal model [43] or by assuming an "encroachment" of the relaxation time of a γ-relaxation by the JG β-relaxation [44]. The inset of Fig. 3 demonstrates that, despite different temperature dependences of the α-relaxation times in both materials, their β-relaxation times coincide at $T < T_g$. This fact was already reported in an earlier work by Minoguchi et al. [31] and found also for other polyols and their mixtures [45]. One may ask how this finding can be brought in line with the concept of a universal JG β-relaxation, which is assumed to be inherent to the glassy state of matter. In [31], based on this finding, hydrogen bond dynamics, being the same in the different polyols, was assumed as the origin of their β-relaxations.

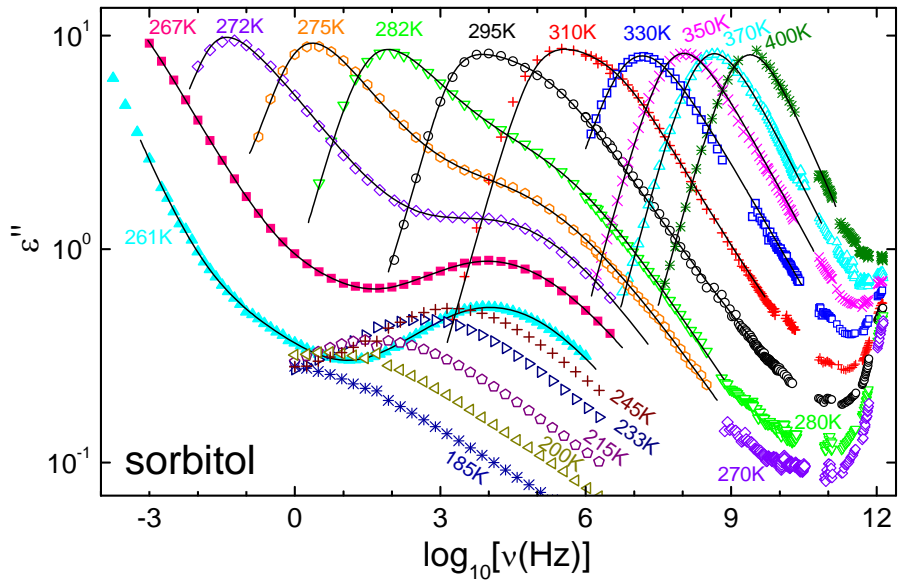

Fig. 2. Broadband dielectric loss spectra of glass forming sorbitol at selected temperatures. The lines are fits with the sum of a Havriliak-Negami and a Cole-Cole function.

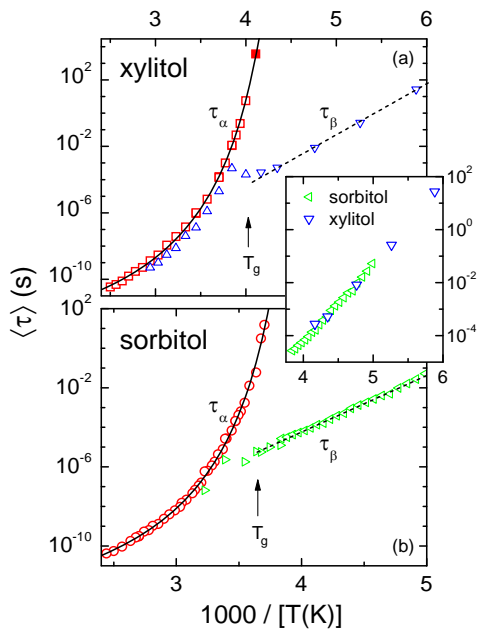

Fig. 3. Temperature dependent α- and β-relaxation times of xylitol (a) and sorbitol (b). For $\tau_\beta$, the upward- and right-pointing triangles denote data that were collected in thermodynamic equilibrium. The solid lines are fits with the Vogel-Fulcher-Tammann law (xylitol: $D = 6.81$, $T_{VF} = 207$ K; sorbitol: $D = 5.17$, $T_{VF} = 233$ K). The dashed lines indicate Arrhenius behavior of $\tau_\beta$ below $T_g$. The inset demonstrates that $\tau_\beta(T)$ of both materials agrees at $T < T_g$.

### 3.2. Fast β-process

Figure 4 presents a magnified view of the high-frequency region of the loss spectra shown in Figs. 1 and 2. Depending on temperature, the left flank of the minimum corresponds to the right flank of the α- or the β-peak or a combination of both (cf. Figs. 1 and 2). It is clear that a minimum must be present in this region because the loss must increase again towards the microscopic excitations, well known to arise in glass-forming materials in the infrared region (e.g., [46,47]). As an example, for one temperature in each material, the dashed line, representing the sum of two power laws, was calculated. The power laws were chosen to match the high-frequency flank of the α- and/or β-peak and the low-frequency flank of the microscopic peak (sometimes termed boson peak) that can be expected to show up at about 2-3 GHz, based on the observations in other glass formers [2,3,14]. Obviously the experimental data cannot be described in such a way. Thus also in these typical type B glass formers an additional fast process seems to be active in this region. Fast processes in glass formers leading to a shallow susceptibility minimum (i.e., loss minimum in the case of dielectric spectra) were predicted by MCT and the experimental results often can be well described within the framework of this theory (see, e.g., [2,3,5,13,15,48]). The fast process is thought to arise from a "rattling" movement of a particle in the transient "cage" formed by its neighbors [5]. Also other approaches were proposed to explain the experimental findings, e.g., within the extended coupling model [7], where a nearly-constant-loss contribution arising from "caged dynamics" explains the shallowness of the minimum.

In view of recent developments of MCT, the finding of a fast process in the minimum region of type B glass formers is a non-trivial result: Recently some efforts have been made to explain the JG β-process and excess wing within MCT [8]. It was shown that the fast β-process in fact can lead to a peak whose spectral form is well approximated by a Cole-Cole function. This peak can be located at relatively low frequency and in this way the MCT may explain the excess wings or the JG β-relaxations showing up in most glass formers. Within this scenario the loss minimum can be expected to arise from a pure superposition of slow β- and microscopic peak (see Fig. 1 of Ref. [49] for a schematic visualization of this situation). However, in case of the two type B glass formers investigated in the present work, the fast process leads to significant contributions in the region of the high-frequency loss minimum and therefore here such a scenario seems unlikely.

Within the idealized version of the MCT, the minimum can be approximated by the sum of two power laws, the von-Schweidler law, $\nu^b$, and the critical law, $\nu^a$ [5]:

$$\varepsilon'' = \frac{\varepsilon_{\min}}{a+b}\left[a\left(\frac{\nu}{\nu_{\min}}\right)^{-b} + b\left(\frac{\nu}{\nu_{\min}}\right)^{a}\right] \quad (1)$$

Here $\nu_{\min}$ and $\varepsilon_{\min}$ denote position and amplitude of the minimum, respectively. The exponents $a$ and $b$ are correlated to each other (i.e., one can be calculated from the other) and should not depend on temperature. Within original MCT, not involving any slow β-relaxations or excess wings, it is reasonable to assume that the exponent $b$ should be identical to the exponent $\beta$ of the high-frequency flank of the α-peak. Already a simple inspection of Figs. 1, 2, and 4 reveals that at high temperatures this indeed is fulfilled for the present cases of xylitol and sorbitol [50]. However, at low temperatures, Figs. 1 and 2 demonstrate that this no longer is the case and the low-frequency wing of the minimum is solely determined by the β-peak. Thus the question arises if in such a case the minimum still may follow the MCT prediction, Eq. (1), i.e., if the von-Schweidler law is unaffected by the transition from the right wing of the α- (or αβ- [50]) to that of the β-peak.

1.5 - 3 decades could be achieved, with somewhat better quality of the fits for xylitol when compared to sorbitol. For low temperatures, $T < T_c$ with $T_c$ the critical temperature of MCT, Eq. (1) no longer is expected to be valid. In addition, only part of the right wing of the minimum can be covered by the fits before the steeper increase towards the microscopic peak sets in. In this respect the situation seems to be similar as in the type A glass former glycerol [3]. Irrespective if the left wing of the minimum is governed by the α- or by the β-peak, the same von-Schweidler exponent $b$ could be used in the fits.

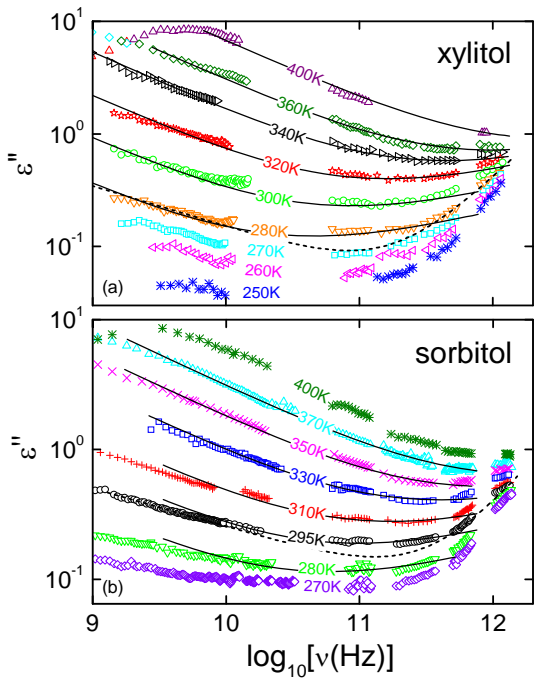

Fig. 4. Frequency dependent loss at selected temperatures of xylitol (a) and sorbitol (b) in the minimum region at high frequencies. The dashed lines demonstrate that a simple superposition of two power laws for the high-frequency flank of the α- and/or β- peak and the low-frequency flank of the microscopic peak is not sufficient to explain the shallow minimum. The solid lines are fits with Eq. (1) with exponents $a$ and $b$ identical for all temperatures ($a = 0.297$ and $b = 0.53$ for xylitol and $a = 0.283$ and $b = 0.49$ for sorbitol).

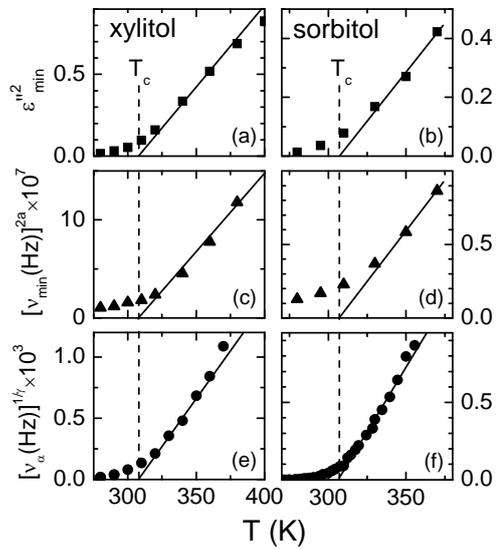

Fig. 5. Temperature dependence of the amplitude $\varepsilon_{min}$ (a, b) and position $\nu_{min}$ (c, d) of the $\varepsilon''(\nu)$-minimum and of the α-relaxation rate $\nu_\alpha$ (e, f) of xylitol and sorbitol. $\varepsilon_{min}$ and $\nu_{min}$ have been taken from the fits with equation (1), shown in Fig. 4. Representations have been chosen that should result in linear behavior according to the predictions of the MCT. The solid lines demonstrate a consistent description of all three quantities with a $T_c$ of 307 K for sorbitol and 308 K for xylitol.

The solid lines in Fig. 4 are fits of the minimum region using Eq. (1) with temperature independent exponents $a$ and $b$, adhering to the correlation of both quantities predicted by MCT [5]. For both materials, especially at high temperatures, a reasonable agreement of fit and experimental data over frequency ranges of

An important prediction of MCT is the critical temperature dependence of α-relaxation time and of the amplitude and frequency position of the minimum [5], namely $\nu_\alpha = 1/(2\pi\tau_\alpha) \propto (T-T_c)^\gamma$, $\nu_{min} \propto (T-T_c)^{1/(2a)}$, and $\varepsilon_{min} \propto (T-T_c)^{1/2}$. Obviously the temperature-dependent shift of the minimum frequency should be directly related to the exponent of the right wing of the minimum. As $\gamma = 1/(2a) + 1/(2b)$ also the shift of the α-peak frequency should be related to the exponents of the high-frequency minimum. In Fig. 5 the temperature dependences of $\nu_\alpha$, $\nu_{min}$, and $\varepsilon_{min}$ are shown. For the ordinates, representations were chosen that should linearize the predicted critical laws with a crossing of the abscissa at $T = T_c$. As demonstrated by the solid lines, for xylitol all three quantities can be consistently described with the same $T_c$. Again, sorbitol shows somewhat less convincing agreement with the theoretical predictions than xylitol. We arrive at $T_c = 308$ K ($= 1.24\ T_g$) for xylitol and at $T_c = 307$ K ($= 1.12\ T_g$) for sorbitol. The linear behavior in Fig. 5 breaks down at low temperatures, close to $T_c$. Partly also at high temperatures deviations show up. This may be explained by the fact that the critical MCT laws should hold for temperatures above but close to $T_c$ only. Overall, the critical laws of MCT in these two materials are as well fulfilled as for most other glass formers reported in literature. In any case, it must be clearly stated that the present evaluation within idealized MCT should be regarded as a first check for consistency of the experimental data with MCT and for obtaining a first estimate of the critical temperature. However, only an analysis within extended versions of MCT can reveal definite information, which, however, is out of the scope of the present work (for examples, see, [51]). It also should be noted that also alternative descriptions of the minimum region may be possible. For example, in Fig. 4 the loss is found to be nearly constant over 1-2 frequency decades in the minimum region, which may be interpreted as an indication of a nearly-constant-loss contribution, which is rationalized, e.g., within the extended coupling model [7].

## 3. Summary and conclusions

In summary, in the present work we have provided dielectric loss spectra of two glass formers with well pronounced JG β-relaxation. The spectra extend up to frequencies of 1.4 THz thus covering the region of the fast β-process. Information on the α- and JG β-relaxation in these systems is provided including some evidence for a third relaxation process in sorbitol arising after prolonged waiting time to achieve thermodynamic equilibrium below $T_g$. In addition, we found evidence for the presence of a fast β-process at GHz - THz in these canonical type B glass formers. A first analysis within idealized MCT leads to critical temperatures of $T_c = 308$ K (xylitol) and $T_c = 307$ K (sorbitol). It seems that the presence of a strong β-peak at low temperatures does not impede the development of the typical critical fast dynamics in these type B glass formers.

## Acknowledgements

This work was partly supported by the Deutsche Forschungsgemeinschaft via research unit FOR 1394.